\documentclass[dtaft]{JHEP3}
\usepackage{bbm}
\usepackage{mathrsfs}
\usepackage{bbm,epsfig}
\title{Quantizing Strings  in de Sitter Space}
\author{Miao Li\\Interdisciplinary Center for Theoretical Study, University of Science and Technology of
China, Hefei, Anhui 230026, China\\
Institute of Theoretical Physics,Academia Sinica, Beijing 100080, China \\
Interdisciplinary Center of Theoretical Studies, Academia Sinica, Beijing 100080, China\\
E-mail: \email{mli@itp.ac.cn}}

\author{Wei Song\\
Institute of Theoretical Physics,Academia Sinica, Beijing 100080, China \\
Interdisciplinary Center for Theoretical Study, University of
Science and Technology of
China, Hefei, Anhui 230026, China\\
 E-mail: \email{wsong@itp.ac.cn}}
\author{Yushu Song\\
Institute of Theoretical Physics,Academia Sinica, Beijing 100080, China \\
Interdisciplinary Center for Theoretical Study, University of
Science and Technology of
China, Hefei, Anhui 230026, China\\
 E-mail: \email{yssong@itp.ac.cn}}
\abstract{We quantize a string in the de Sitter background, and we
find that the mass spectrum is modified by a term which is quadratic
in oscillating numbers, and also proportional to the square of the
Hubble constant.} \keywords{String theory and cosmic strings
, Bosonic Strings }
\preprint{USTC-ICTS-07-01}
\def\be{\begin{equation}}
\def\ee{\end{equation}}
\def\ba{\begin{eqnarray}}
\def\ea{\end{eqnarray}}
\def\n{\nonumber}
\def\p{\partial}

\def\rt{e^{2Ht}}
\def\ep{h_{\sigma\sigma}}
\def\iep{h^{-1}_{\sigma\sigma}}

\def\lfr{{a^i_m(t)\over\sqrt{2\,|\lambda_m(t)|}}\,e^{-i\int^t
du\,\lambda_m(u)}\,e^{im\sigma}}
\def\rfr{{\tilde{a}^i_m(t)\over\sqrt{2|\lambda_m(t)|}}\,e^{-i\int^t
du\,\lambda_m(u)}\,e^{-im\sigma}}
\def\n{\nonumber}

\def\x{{\dot{\omega}\over2\omega}}
\begin{document}

\section{Introduction}

The progress of the string program as a theory of quantum gravity
and other interactions is currently impeded by a fundamental
difficulty, namely we do not know how to formulate string theory in
a time-dependent background in general, and how to understand many
issues related to cosmology such as the origin of our universe and
the nature of dark energy in particular. This baffling situation
leads to lots of debates about whether string theory is the correct
theory of nature, and whether string theory has any predictive power
if there exists a vast landscape of meta-stable vacua. It goes
without saying that string theory has been tremendously successful
in resolving some of deeper conceptual problems such as whether
gravity is compatible with quantum mechanics, but only in some
unrealistic backgrounds such as a flat background and an anti-de
Sitter background. In some cases, we even have a non-perturbative
formulation, for instance, a CFT is a non-perturbative theory in the
AdS/CFT duality. Nevertheless, until we have a theory for
time-evolving backgrounds, string theory can not claim to be the
theory of our universe.

We shall not try to attack the ultimately difficult problem of
formulating string theory in a general or even an ``on-shell"
time-dependent background in this note. Our purpose is rather
pragmatic, we will try to work out part of string quantization in a
de Sitter background, with applications to inflation as well as to a
later universe dominated by dark energy in mind. For instance, we
would like to know how different the spectrum of a string in the de
Sitter space is from that in the flat spacetime, whether this
spectrum enables string production during inflation. If not, whether
strings are created at the end of inflation when the Hubble constant
undergoes transition from a constant to a decreasing function. In
the later universe such as the current epoch, our universe is again
dominated by energy of almost constant density. Although a cosmic
string, if exists, is largely governed by classic dynamics, it is
certainly of interest to know whether its spectrum is modified in
some extremal limit.

The answers to the above questions seem to be yes. As we shall see,
the dynamic equation for the field corresponding to a fixed state
contains a new term induced by ``string mode creation" (to be
explained shortly). This term depends on the Hubble constant, thus
it renders string creation possible in the end of inflation. This
term begins to be comparable to the usual term in the ``mass"
spectrum when the oscillation numbers are large enough. We put mass
into quotation marks since there is no notion of mass in a de Sitter
space.

Note that, we will exclusively deal with ``small" string states in
this paper, by a small string we mean that the string modes are
mostly oscillating. It is known that cosmic strings are ``long"
strings, namely the dominating modes are not oscillating modes so
that the major part of the string co-moves with the expansion of the
universe.

There is a series of papers on first quantizing string in de Sitter
space by de Vega and S$\acute{\hbox{a}}$nchez and their
collaborators, see for instance \cite{sanchez}-\cite{sanchezf}. In
their work, they fix all the degrees of freedom of the world sheet
metric, and find an exact classical solution. To quantize, they
propose two methods, one is to quantize the fluctuation around an
exact solution of the center of mass \cite{sanchez}, and the other
is to propose a quantization condition semiclassically \cite{semi}.

In this note, we propose a new approach. We leave one degree of the
world sheet metric unfixed, and then eliminate it by a constraint.
To the leading order, our result agrees with that of de Vega and
S$\acute{\hbox{a}}$nchez. However, there is a subtle difference: our
method is approximate in choosing the gauge. Despite of this, the
answer is exact once the gauge is chosen. The most important
consequence of our result is that the mode creation and annihilation
operators are still time-dependent even after diagonalization. Thus,
a state created by these operators is itself time-dependent, and
this will have some effects in the dynamic equation for the
corresponding field.

We will present our approach in the next section and carry out the
first quantization in Sect.3. We will discuss possible applications
to inflation and obtain some conclusions in the last section. A
discussion on the mode creation on a string is left in the appendix.

\section{Action and gauge choice}

We start with the Polyakov action in a general background
 \be \label{paction}
S=-\frac{1}{4\pi \alpha'}\int
d\tau\,d\sigma\,\sqrt{-h}\,h^{ab}\,\partial_a X^\mu\,\p_b
X^\nu\,G_{\mu\nu}(X)\ ,  \ee
 where $h=\det h_{ab}$ ($a$, $b$ run over
values $(\tau,\sigma)$.), $0\leq \sigma \leq 2\pi$, and $G_{\mu\nu}$
is the string frame metric. Here we suppose that the dilaton is
constant, and $G_{\mu\nu}$ ($\mu,\nu$ run over 0, 1, 2, 3. ) is the
metric of de Sitter space in comoving frame
\be\label{comoving}ds^2=-dt^2\,+\,e^{2Ht}\,(dx^i)^2\ ,\,i=1,2, 3\ .
\ee Of course, if we naively take (\ref{paction})\ as the whole
story, then in the background (\ref{comoving}) a quantum string is
not well-defined, since we do not have a two dimensional conformal
field theory. We shall assume that there is a hidden sector making
the whole world-sheet action conformally invariant (as an example,
in a merely AdS space, the bosonic string action is not conformally
invariant, and we need another sector from a sphere as well some
other terms due to flux and fermionic degrees. As for a de Sitter
space, we assume a scenario such as KKLT compactification making the
story complete).

By means of the classical world-sheet symmetry, we can set
determinant of the world-sheet metric to $-1$ after choosing the
temporal gauge and diagonalizing world-sheet metric. Following this
strategy, there is only one component of the world sheet metric
$\ep$ left unfixed, which is non-dynamical. We assume that $\ep$
depends only on time, then the target space coordinates $X^i$ can be
solved in terms of $\ep$. When we quantize the field $X^i$, $\ep$
will be promoted to an operator. In order to obtain the on-shell
condition, we then impose the constraint from the variation of $\ep$
on physical states. In more detail, the constraint is the integral
of the variation of $\ep$ due to our assumption.

Now we perform the steps summarized above. By choosing a proper
gauge, we can fix the redundancies in the Polyakov
action, and make the equations of motion simple. Set\\
\be\label{gauge} \tau=t\ ,\,h^{\tau\sigma}=0\ ,\,-h=1\ . \ee where
$t$ is the comoving time. Under this gauge choice, the action
becomes \be\label{actone} S=\frac{1}{4\pi \alpha'}\int
dt\,d\sigma\{-h_{\sigma\sigma}\,+\rt\,[h_{\sigma\sigma}\,(\p_t
X^i)^2-h^{-1}_{\sigma\sigma}\,(\p_\sigma X^i)^2]\}\ . \ee

There are two independent constraints due to functional variation of
$h^{ab}$, that is,

 \ba\label{constrone}&&\frac{1}{4\pi
\alpha'}\,\rt\,\p_t
X^i(t,\sigma)\,\p_\sigma X^i(t,\sigma)=0\ ,\\
&&\frac{1}{8\pi
\alpha'}\{-\ep^2(t,\sigma)+\rt\,[\ep^2(t,\sigma)(\p_t
X^i(t,\sigma))^2+(\p_\sigma X^i(t,\sigma))^2]\}\ =0\
.\label{constrtwo} \ea

The equations of motion corresponding to the functional variation of $X^{\mu}$ are
\ba\label{eom} &&\p_t(\rt\,\ep(t,\sigma)\,\p_t
X^i(t,\sigma))-\rt\,\p_\sigma(\iep(t,\sigma)\,\p_\sigma
X^i(t,\sigma))=0\ ,
\\&&\p_t\ep(t,\sigma)+H\,e^{2Ht}\, [\ep(t,\sigma)
(\p_tX^i(t,\sigma))^2-\ep^{-1}(t,\sigma)(\p_\sigma
X^i(t,\sigma))^2]=0\ . \label{constrthree}\ea

The second equation of motion (\ref{constrthree}) comes from the
functional variation of $X^0$, which is non-dynamical according to
our gauge choice. This becomes another constraint, but fortunately
it can be derived from the other three equations. So at last we get
two constraints and three equations of motion (each for one spatial
coordinate).

The conjugate momentum of $X^i(t,\sigma)$ is given by
$\Pi_i(t,\sigma)=\frac{1}{2\pi
\alpha'}\,\rt\,h_{\sigma\sigma}(t,\sigma)\,\p_t X^i(t,\sigma)\ .$
The Hamiltonian is then \ba\label{hamil} E&&=\frac{\rt\,}{4\pi
\alpha'}\int
d\sigma\,[\frac{\ep(t,\sigma)}{\rt\,}+\ep(t,\sigma)(\p_t
X^i(t,\sigma))^2+\,h^{-1}_{\sigma\sigma}(t,\sigma)\,(\p_\sigma
X^i(t,\sigma))^2]\\\
&&\simeq \frac{1}{2\pi \alpha'}\int
d\sigma\,\ep(t,\sigma)\label{density} \ . \ea In the last step we
used (\ref{constrtwo}), which is satisfied only by physical states.
We use the symbol $\simeq$ instead of $=$ to show that the equality
is satisfied only by  physical states. Hence $\frac{1}{2\pi
\alpha'}\int d\sigma\ep$ is in fact just the energy of a physical
state with respect to comoving time. For simplicity, we will set
$\alpha'=1$ hereafter. $\ep$ is non-dynamical, and is determined by
$X^i(t,\sigma)$ through (\ref{constrtwo}). If we eliminate $\ep$,
(\ref{eom}) will become nonlinear equations which is hard to solve.
Instead, we will first treat $\ep$ as an independent variable, solve
the equations of motion for $X^i(t,\sigma)$ in terms of $\ep$ and
then fix $\ep$ by (\ref{constrtwo}). In order to solve the equation
of motion (\ref{eom}), we make an assumption that
\be\label{ansatz}\ep(t,\sigma)=\omega(t) . \ee This is the only
approximation we execute in this note. For strings oscillating fast,
$\omega (t)$ may be viewed as an average of $\ep(\sigma, t)$ along
$\sigma$.

Hereafter we will just write $\omega$ instead of $\omega(t)$ for
simplicity, but keep in mind that $\omega$ is in fact a function of
time. Also, we caution that upon quantization, $X^i$ become operators, so does
$\omega$, namely $\omega$ is not to be viewed as a usual function.

For physical states, $E\simeq\omega$ is the energy of the string,
the equations of motion become
\be\p_t(\eta^{-2}\,\p_tX^i(t,\sigma))-\omega^{-2}\,\eta^{-2}\,\p^2_\sigma
X^i(t,\sigma)=0\ ,\ee where $\eta=\frac{1}{ e^{Ht} \sqrt{\omega}}$.
A general solution is \ba\label{solut} \n X^i(t,\sigma)&&=x_0+\int
^t du\,\eta^2(u)\,p^i+\sum_{m\in Z/\{0\}}\eta(t)[\lfr\\&&+\rfr]\
,\\\dot{a}^i_m(t)&&={\dot{\lambda}_m(t)\over2\lambda_m(t)}\,\tilde{a}^i_{-m}(t)
\,e^{2i\int^t du \lambda_m(u)},\,
\dot{\tilde{a}}^i_m(t)={\dot{\lambda}_m(t)\over2\lambda_m(t)}\,a^i_{-m}(t)\,
\,e^{2i\int^t du \lambda_m(u)}\ .\label{chag}\ea where we have
defined$\lambda_m=sgn(m)\sqrt{{m^2\over\omega^2}-\eta\p^2_t(\eta^{-1})}=
sgn(m)\sqrt{{m^2\over\omega^2}-(H+{\dot{\omega}\over2\omega})^2-\p_t
({\dot{\omega}\over2\omega})}$, where the function $sgn(m)=1$ for
$m>0$, and $sgn(m)=-1$ for $m<0$ . We will work in the situation
that $\lambda_m$s remain real, which means that the string is
oscillating in time. This is to be dubbed as a small string, since
it is not stretched too much with the expansion of the universe.

The real condition for $\lambda_m$ is that
${m^2\over\omega^2}-(H+{\dot{\omega}\over2\omega})^2-\p_t
({\dot{\omega}\over2\omega})>0$. The Hermiticy of $X^i(t,\sigma)$
requires that $(a_m^i)^\dag=a^i_{-m}$ and
$(\tilde{a}^i_m)^\dag=\tilde{a}^i_{-m}$. Thus the conjugate momentum of $X^i$
becomes \ba\nonumber\Pi^i(t,\sigma)&&={1\over 2\pi}\{p^i+\sum_{m\in
Z/\{0\}}[{\dot{\eta}(t)\over\eta(t)^2}-{i\lambda_m(t)\over\eta(t)}]
\,[\lfr\\
\n&&+\rfr]\}\ .\ea
To quantize, impose the equal time canonical
commutation relations
\ba\label{commut}[X^i(t,\sigma),X^j(t,\sigma')]=[\Pi_i(t,\sigma),\Pi_j(t,\sigma')]=0\ ,\\
\,[X^i(t,\sigma),\Pi_j(t,\sigma')]=i\,\delta^i_j\,\delta(\sigma-\sigma')\
.\ea which are equivalent to imposing the commutation relations \ba
&&[x^i,x^j]=[p^i,p^j]=0\
,\,[x^i,p^j]=i\delta^{ij}\,\label{cmf}\\&&[a_m^i,\tilde{a}_n^j]=0\
,\, [a_m^i, a_n^j]={m\over|m|}\delta^{ij}\delta_{m,-n}\label{cmth}\
.\ea

Note that $a^i_m$ and $\tilde{a}^i_m$ depend on $\omega$ implicitly.
Thus when we impose (\ref{cmth}), we have also promote $\omega$ to
be an operator, which commutes with other operators including
$\dot{\omega}$. To simplify the notations, we will not distinguish
operators and functions explicitly except that  $\omega$, $p$ and
$N$ are assumed to be operators when constructing states.

\section{Quantization}

We now study constraints (\ref{constrone}), (\ref{constrtwo})
 and (\ref{constrthree}). (\ref{constrthree}) can be derived from
(\ref{constrone}), (\ref{constrtwo}) together with the equation of
motion (\ref{eom}). Thus only (\ref{constrone}) and
(\ref{constrtwo}) need to be considered. According to our assumption
$\ep(t,\sigma)=\omega(t)$, the most important parts of these
constraints are  their average over $\sigma$. Thus the constrains
become,
 \ba \n P&&\equiv \int {d\sigma\over4\pi}\,\rt\,\p_t
X^i(t,\sigma)\,\p_\sigma
X^i(t,\sigma)\\&&=\sum_{m>0}{m\over2\omega}\,\{a^i_{-m}\,a^i_{m}-\tilde{a}^i_{-m}\,\tilde{a}^i_m
\}\simeq0\ , \label{level}\ea
and
\ba\label{hmtlnnc}\mathscr{H}&&\equiv \int
{d\sigma\over8\pi}\,\{\ep^2\,[-1+\rt\,(\p_t X^i(t,\sigma))^2]+\rt\,(\p_\sigma X^i(t,\sigma))^2\}\\
\n&&=-{\omega^2\over4}+{(p^i)^2\over4\,\rt}+\sum_{m>0,i}{\omega\over4\lambda_m}\,
[({\dot{\eta}\over\eta})^2+\lambda_m^2+{m^2\over\omega^2}]\,
(a^i_{-m}\,a^i_m+\tilde{a}^i_{-m}\,\tilde{a}^i_m+1)\\
\n&&+{\omega\over4\lambda_m}e^{-2i\int^t du
\lambda_m(u)}\,[({\dot{\eta}\over\eta}-i\lambda_m)^2+{m^2\over\omega^2}]\,
a^i_m\,\tilde{a}^i_m \\&&+{\omega\over4\lambda_m}e^{2i\int^t du
\lambda_m(u)}\,[({\dot{\eta}\over\eta}+i\lambda_m)^2+{m^2\over\omega^2}]\,
a^i_{-m}\,\tilde{a}^i_{-m}\simeq0\ .\ea

There are also infinitely many constraints corresponding to positve
modes of (\ref{constrone}), (\ref{constrtwo}) in the Fourier expansion
in terms of $\sigma$. Note that our temporal gauge choice only eliminate
the temporal degrees of freedom. These infinitely many constraints, whose
analog in flat spacetime are the conditions $L_n=\tilde{L}_n=0$, will further
 eliminate the unphysical degrees of freedom due to the
longitudinal excitations. In this note we will not discuss them in detail,
and will just focus on the mass shell condition, whose analog in flat
spacetime is $L_0=0$. One may ask that why there are still two set of constraints
while only the longitudinal degrees of freedom need to be eliminated.
Are our constraints too strong? The answer is no, because we have used the
assumption that $h_{\sigma\sigma}$ is independent of $\sigma$, which imposes another
constraint. If we re-introduce $h_{\sigma\sigma}$ as an arbitrary function of
$\sigma$, the counting of degrees of freedom will be correct.

Define the occupation number $n^i_{m}=a^i_{-m}\,a^i_m$, and the
level $n=\sum_{m,i}m\,n^i_{m} $, and similarly for $\tilde{n}$. Then
the vanishing of (\ref{level}) on physical states is just the level
matching condition $n=\tilde{n}$. Note that the condition
(\ref{level}) is also the translational invariance condition along
$\sigma$. We can make a linear transformation to define another set
of creation and annihilation operators,
\ba\label{chbassis}&&A_m^i=\alpha_m\,a^i_m+\beta_m\,\tilde{a}^i_{-m}
,\,\tilde{A}^i=\tilde{\alpha}_m\,\tilde{a}^i_m+\tilde{\beta}_m\,a^i_{-m}\
,\\&&|\alpha_m|^2-|\beta_m|^2=|\tilde{\alpha}_m|^2-|\tilde{\beta}_m|^2=1\
,\label{cmo}\\&&\alpha_m\beta_{-m}=\tilde{\alpha}_m\tilde{\beta}_{-m}\
.\label{cmt}\ea The conditions (\ref{cmo}) and (\ref{cmt}) ensure
that the new operators satisfy the same commutation relation as
(\ref{cmth}). The above transformaion may be viewed as Bogoliubov
transformation on the world-sheet. The most general form is \ba
&&\alpha_m=\cosh(\gamma_m)\,e^{i\delta_m+i\phi_m}, \tilde{\alpha}_m=
\cosh(\gamma_m)\,e^{i\delta_m+i\psi_m}
,\,\\&&\beta_m=\sinh(\gamma_m)\,e^{i\phi_m},\, \quad \,
\tilde{\beta}_m=\sinh(\gamma_m)\,e^{i\psi_m}\ .\label{commcdt}\ea
where $\gamma_m, \phi_m, \psi_m$ and $\delta_m$ are real. Constraint
($\ref{level}$) remains  the level matching condition, but now in
terms of the new occupation number operator $N=\tilde{N}$, defined
by $N=\sum_{i,m}mN^i_m$, and $N^i_m\equiv A^i_{-m}\,A^i_m $, and
similarly for $\tilde{N}$.

 We now choose the parameters
$\gamma_m$ and $\delta_m$ properly to diagonalize the constraint
(\ref{hmtlnnc}). The conditions are
\ba&&\cosh^2(\gamma_m)
+\sinh^2(\gamma_m)={\omega\over2m\,\lambda_m}\,
[({\dot{\eta}\over\eta})^2+\lambda_m^2+{m^2\over\omega^2}]\
,\\
&&\sinh(2\gamma_m)={\omega\over2m\,\lambda_m}\,
[({\dot{\eta}\over\eta}-i\lambda_m)^2+{m^2\over\omega^2}]\,e^{-2i\int^t
du\, \lambda_m(u)-i\delta_m}\,\ea where $\delta_m$ is chosen to make
$\sinh(2\gamma_m)$ real, and this can be done.

In terms of the new creation and annihilation operators
\ba\mathscr{H}&&=-{\omega^2\over4}+{({p^i})^2\over4e^{2Ht}}
+\sum_{m>0,i}\frac{m}{2}(N^i_m+\tilde{N}^i_{m}+1)\
\\\n&&\simeq-{\omega^2\over4}+{(p^i)^2\over4e^{2Ht}}
+N+\frac{E_0}{2}\simeq0\ ,\,E_0=-{1\over 4}\ \label{ccon}.\ea
The fact that we need to form a Bogoliubov transformation implies that
if we start with a state constructed by the original operators $a^i_m$,
there will be mode creation along the string at a later time.

We now discuss implication for dynamics of fields viewed as coefficients in the
expansion of a general string state.

Since $\p_{i}$ is a Killing vector, $p^i$ is conserved, and we will
just work in momentum representation. Any physical state should
satisfy the constraint $\mathscr{H}=0$, and can be expanded in terms
of common eigenstates of the occupation number operator $N^i_m$s,
$\tilde{N}^i_m$s, $p^i$ as well as $\omega$. That is,\be|\phi>=
\sum_{N^1_1,\tilde{N}^1_1,.....N^i_m,\tilde{N}^i_m,......}|N^1_1,\tilde{N}^1_1,.....N^i_m,
\tilde{N}^i_m,.......\omega,p^i>\phi(N^i_m,\tilde{N}^i_m,\omega,p^i)\
,\ee where the eigenvalues labeling the states must satisfy the
relation $-\omega^2+{(p^i)^2\over e^{2Ht}}+2(2N+E_0)=0$, and
$N=\tilde{N}$.

When writing down the action for the above general state, it is
important to keep in mind that the inner product involves an
integral over space so there is a nontrivial Hermticity condition.
For instance, we consider a scalar particle with mass $m$, whose
wave function must satisfy
\be(\square-m^2)\phi(x)=[-\p_t^2-3H\p_t+e^{-2Ht}(\p^{i})^2-m^2]\phi(x)=0\
.\ee Written in momentum representation, the last two terms are
$-e^{-2Ht}\ (p^i)^2-m^2=-E^2$, where $E$ is the comoving energy.

In string field theory, we assume that the action of string state
has the form of \begin{equation}   S=\int
dt<\phi|\partial_t^2+3H\partial_t+E^2+\lambda\mathscr{H}|\phi>\end{equation}
where $\lambda$ is the Lagrangian multiplier. In writing down this action,
we have only considered the mass shell condition and have omitted other constraints corresponding to the positive
Fourier modes. To treat the problem more rigorously, one should introduce more Lagrangian multipliers.
 From this action, we
can easily get the evolution equation of the string state. By
variation of $|\phi>$, we have
\be(\p_t^2+3H\p_t+E^2)|\phi>=[\p_t^2+3H\p_t+e^{-2Ht}\ (p^i)^2
+2(N+\tilde{N}+E_0)]|\phi>=0\ .\ee This equation contains an
unphysical component which is to be discarded due to the fact that
in the action the inner production automatically projects out the
unphysical component by imposing the constraint
$\mathscr{H}|\phi>=0$.

We explain  some subtleties in deriving
this second order equation. Because of the non-flat
metric, the measure of the integral volume is
$d\vec{x}^3\sqrt{-G}=d\vec{x}^3e^{3Ht}$, so the inner product should
be defined as $\int d^3xe^{3Ht}\phi(x)^*\phi(x)$. With this inner product,
$\p^2_t$ is not Hermitian. To get a Hermitian operator, we should
replace $\p^2_t$ with $\p^2_t+3H\p_t$. There is no
addition term caused by polarization indices since all the creation operators are
properly normalized.

Now, different excitation modes of the string correspond to
different particles in spacetime, thus the coefficient
$\phi(N^i_m,\tilde{N}^i_m,\omega,p^i)$ is the wave function of
single particle. From now on, we will use notation
$|N^i_m,\tilde{N}^i_m,\omega,p^i>$ instead of
$|N^1_1,\tilde{N}^1_1,.....N^i_m, \tilde{N}^i_m,.......\omega,p^i>$
for simplicity. Since the basis $|N^i_m,\tilde{N}^i_m,\omega,p^i>$
evolves with time, this dependence on time will be transmitted into
the equation of motion for $\phi$ through the action of
$\partial_t^2+3H\partial_t$ . Using the differential equation for
$a_m^i$ and $\tilde{a}^i_m$, (\ref{chag}),
  and the
definition of $A^i_m$, $\tilde{A}^i_m$, we have
\ba\dot{A}_{-m}^i&&=c_m\,\tilde{A}^i_m+d_m\,A^i_{-m},\,
\dot{\tilde{A}}_{-m}^i=\tilde{c}_m\,A^i_m+\tilde{d}_m\,\tilde{A}^i_{-m}\ ,\\
\n
c_m&&=e^{i(\phi-\psi)}\{\alpha^*_m\dot{\beta}^*_m-\beta^*_m\dot{\alpha}^*_m
+[{\alpha^*}^2e^{-2i\int^t du \lambda_m(u)}-{\beta^*}^2e^{2i\int^t
du
\lambda_m(u)}]{\dot{\lambda}_m\over2\lambda_m}\}\\&&=e^{(-i\delta-i\phi-i\psi)}H{\p_t\x
-i{2m\over\omega}(H+\x)\over\sqrt{({2m\over\omega})^2
(H+\x)^2+[\p_t(\x)]^2}}\ ,
\\\n
\tilde{c}_m&&=e^{i(\psi-\phi)}\{\tilde\alpha^*_m\dot{\tilde
\beta}^*_m-\tilde\beta^*_m\dot{\tilde\alpha}^*_m
+[({\tilde\alpha}^*)^2e^{-2i\int^t du
\lambda_m(u)}-({\tilde\beta}^*)^2e^{2i\int^t du
\lambda_m(u)}]{\dot{\lambda}_m\over2\lambda_m}\}\\&&=e^{(-i\delta-i\phi-i\psi)}H{\p_t\x
-i{2m\over\omega}(H+\x)\over\sqrt{({2m\over\omega})^2
(H+\x)^2+[\p_t(\x)]^2}}\ ,
\\\n
d_m&&=e^{i(\phi-\psi)}\{\tilde{\alpha}_m\dot{\alpha}^*_m-\tilde{\beta}_m\dot{\beta}^*_m
+{\dot{\lambda_m}\over2\lambda_m}[\tilde{\alpha}_m\beta^*_me^{2i\int^t
du\, \lambda_m(u)}-\alpha^*_m\tilde{\beta}_me^{-2i\int^t du\,
\lambda_m(u)}]\}\\
\n&&=i{H+\x\over2\lambda_m^2({2m^2\over\omega^2}
-\p_t\x-{2m\over\omega}\lambda_m)} \{{4Hm\over\omega}\lambda_m^2
+\lambda_m[\p_t^2\x+2(H+\x)\p_t\x+{4m^2\over\omega^2}\x]\}\\
&&-i\dot{\phi}\ ,\\\n
\tilde{d}_m&&=e^{i(\psi-\phi)}\{\alpha_m\dot{\tilde{\alpha}}^*_m
-\beta_m\dot{\tilde{\beta}}^*_m
+{\dot{\lambda_m}\over2\lambda_m}[\alpha_m\tilde{\beta}^*_m
e^{2i\int^tdu\, \lambda_m(u)}-\tilde{\alpha}^*_m\beta_m
e^{-2i\int^tdu\,\lambda_m(u)}]\}\\\n&&=i{H+\x\over2\lambda_m^2({2m^2\over\omega^2}
-\p_t\x-{2m\over\omega}\lambda_m)} \{{4Hm\over\omega}\lambda_m^2
+\lambda_m[\p_t^2\x+2(H+\x)\p_t\x+{4m^2\over\omega^2}\x]\}\\\n
&&-i\dot{\psi} \ .\ea  Note that $|c_m|=H$. We can choose $\phi$ and
$\psi$ to set $d_m=\tilde{d}_m=0$, and then the phase of $c_m$ is
also fixed.

 The eigenstates of $N^i_m$ and $\tilde{N}^i_m$ are just
$\Pi_{m,i}(A^i_{-m})^{N^i_m}(\tilde{A}^i_{-m})^{\tilde{N}^i_m}\,|\Omega,\omega>$,
where $|\Omega,\omega>$ is defined as $A_m^i|\Omega,\omega>=0,$
$\tilde{A}^i_m|\Omega,\omega>=0$ for all $m,i$. From
$\dot{A}^i_m\,|\Omega,\omega>+A^i_m\,\p_t|\Omega,\omega>=0$, we get
\be\p_t|\Omega,\omega>=-\sum_{m,i}c_m^*\,A^i_{-m}\,\tilde{A}^i_{-m}|\Omega,\omega>\
,\ee and\ba
&&\p_t|N^k_n,\tilde{N}^k_n,\omega>=\sum_{m,i}[c_m\,A^i_m\,\tilde{A}^i_m-
c^*_m\,A^i_{-m}\,\tilde{A}^i_{-m}]|N^k_n,\tilde{N}^k_n,\omega>\
,\\&&\p_t^2|N^k_n,\tilde{N}^k_n,\omega>=-\sum_{m,i}|c_m|^2\,
(1+2N^i_{m}\,\tilde{N}^i_{m}+N^i_{m}+\tilde{N}^i_{m})|N^k_n,\tilde{N}^k_n,\omega>\
,\\&&+\{-\sum_{m,i,l,j}2\,c_m\,c^*_l\,A_{-l}^i\,\tilde{A}^i_{-l}\,A^j_{m}\,
\tilde{A}^j_{m}+\sum_{m,i}\dot{c}_m\,A^i_{m}\,\tilde{A}^i_{m}
-\sum_{m,i}\dot{c}^*_m\,A^i_{-m}\,\tilde{A}^i_{-m}\\
&&+\sum_{m,i,l,j}[c_l\,c_m\,A_{l}^i\,\tilde{A}^i_{l}\,A^j_{m}\,
\tilde{A}^j_{m}+c^*_l\,c_m^*\,A_{-l}^i\,\tilde{A}^i_{-l}\,A^j_{-m}\,
\tilde{A}^j_{-m}]\}|N^k_n,\tilde{N}^k_n,\omega>\ ,\\
&&\dot{\omega}|N^k_n,\tilde{N}^k_n,\omega>=-{Hp^2e^{-2Ht}\over\omega}
|N^k_n,\tilde{N}^k_n,\omega>\ ,\\
&&\dot{N}+\dot{\tilde{N}}=\sum_{m,i}2m(c_m\,A^i_m\,\tilde{A}^i_m+
c^*_m\,A^i_{-m}\,\tilde{A}^i_{-m})\ .\ea From the explicit
expression above, we see that the off diagonal parts of
$\p_t|N,\tilde{N},\omega>$ and $\p_t^2|N^i_m,\tilde{N}^i_m,\omega>$
are unphysical if $|N^i_m,\tilde{N}^i_m,\omega>$ is physical. So the
off diagonal part is orthogonal to physical states.

Just considering the physical part of the following equation
\ba\n&&\{\p_t^2+3H\p_t+(p^i)^2\,e^{-2Ht}+4N+2E_0\}|\phi>=0
 \ ,\ea
we have
\ba\label{swe}\n&&\{\p_t^2+3H\p_t+(p^i)^2\,e^{-2Ht}+4N+2E_0\\&&-\sum_{m,i}H^2\,
(1+2N^i_{m}\,\tilde{N}^i_{m}+N^i_{m}+\tilde{N}^i_{m})\}
\phi(N^k_n,\tilde{N}^k_n,\vec{p})=0
 \ .\label{fneq}\ea

The above equation is the main result of this note. We see that in
addition to the term $4N$, there is an additional term which is
quadratic in creation numbers with a prefactor $H^2$. This term
could become comparable to the linear term $N$ for a fixed $H^2$
(measured in the string unit since we have set $\alpha'=1$).
Restoring the string scale $M_s$, we find that the new term is
comparable to the old term $M_s^2N$ when $N\sim M_s^2/H^2$. This is
of course a large number during inflation because $H$ is much
smaller than the string scale, if we hope that an effective field
theory is valid.

This quadratic term is negative when viewed as a contribution to the
mass squared $m^2$, thus it appears possible to have an effective
negative mass squared if the quadratic term becomes dominating. This
will never become possible, since in our approach so far we have
assumed real $\lambda_m$ in the mode expansion (\ref{chag}), and we
are dealing with ``small" strings mostly oscillating in time. It can
be checked that as long as the real condition on $\lambda_m$ is met,
the quadratic term in (\ref{fneq}) will never make the mass squared
negative. Nevertheless, the fact that this term reduces the mass
squared comes as a surprise. We may just imagine that this is a
quantum effect for a highly excited state due to the mode creation
on the string. We leave a discussion on mode creation to the
appendix.

\section{Comparison with earlier results}

The spectrum we get is \be \alpha'M^2=4N+2E_0-\sum_{m,i}H^2\,
(1+2N^i_{m}\,\tilde{N}^i_{m}+N^i_{m}+\tilde{N}^i_{m}),\ee which can
be read from (\ref{swe}). The previous result obtained by the method
of \cite{sanchez} is
\be\alpha'M^2=24\sum_{n>0}{2n^2-H^2M^2\alpha'^2\over\sqrt{n^2-H^2M^2\alpha'^2}}
+2N{2-H^2M^2\alpha'^2\over\sqrt{1-H^2M^2\alpha'^2}}\ ,\ee
 where we cite the formula in the form appearing in \cite{semi},
 and the dimensionality of the spacetime is 25 there.
We can see that when $H=0,$ both results return to the flat
spacetime spectrum, $4N+\hbox{zero point energy}$. The difference
between the zero point energy is due to the different
dimensionality. When $H\neq0$, both results have a term proportional
to $H^2$, but the coefficients are different. What's more, our
result shows that the spectrum depends not only on the level $N$,
but also on the specific excitation. This difference may due to the
gauge choice of the two approaches. As we have mentioned before, we
fix the worldsheet time $\tau$, set the determinant of the
worldsheet metric to be one, set one component of the worldsheet
metric $h^{12}=0$, and leave another worldsheet metric component
$h^{\sigma\sigma}$ unfixed. To set the determinant of the worldsheet
metric to be one, we have used the classical conformal symmetry of
the action, which does not exist in the whole theory. While in the
earlier approach, for example, \cite{sanchez}, they choose the
conformal gauge and leave the worldsheet coordinate $\tau$ unfixed.
Since there is no conformal symmetry, there is no guarantee that we
should get the same result. The other approach previously developed
in \cite{semi}, namely, the semiclassical quantization, also take
the same conformal gauge, and make a circular string ansatz, and
then they propose a quantization condition. The spectrum is
$\alpha'm^2\approx5.9n, n\in N_0$ in \cite{semi}.  In \cite{semi},
the authors compare this result with that obtained in \cite{sanchez}
by calculating the maximum excitation number of string states in de
Sitter spacetime. In \cite{sanchez}, the maximum number of a single
excitation is $N_{max}={0.15\over H^2\alpha'}$, which obtained in
\cite{semi} is $N_{max}={0.17\over H^2\alpha'}$. In our approach,
the condition for the state to oscillate is $\lambda^2_{m}\geq0$.
Together with (\ref{ccon}), we will get roughly $N_{max}={0.25\over
H^2\alpha'}$ for all the oscillating modes to have real frequency.
Our result is slightly larger than the previous results. One
possibility is that our solution is more general than the previous
ones. For instance, in \cite{semi}, only circular solutions are
considered, and in \cite{sanchez}, only expansion around an exact
solution is considered. While here in our paper, only one
approximation is made, namely, the worldsheet metric
$h^{\sigma\sigma}$ does not depend on $\sigma$. Thus we might have
found more solutions.

\section{Discussions and conclusion}

In this paper, we first quantized a general oscillating string in a
de Sitter space , with the only approximation that
$h_{\sigma\sigma}$ depends only on time. This quantity becomes the
energy density along the string after we impose constraint condition
on it. So our approximation amounts to averaging the energy density
along the string. Aside from this, our treatment is exact.

Apparently, our main result (\ref{fneq}) differs from the old result
by a negative contribution to the mass squared of the string. This
term which is quadratic in oscillation numbers increases quickly
when we consider more highly excited states. Applying this result to
inflation, there is virtually no physical effect during inflation
except for modification of the mass spectrum, for this new term
depends only on the Hubble constant thus remains a constant for a
given state. It is a well-known result that for a particle of
constant mass, there is no particle production during inflation.
Although we have worked with a constant Hubble constant, some of our
results can be generalized to a non-constant Hubble constant. We
expect string creation in the end of inflation can be induced by
this new term , for  the Hubble constant is no longer a constant in
this short reheating period . A similar effect is discussed in
\cite{gubsero} and \cite{gubsert} except for ad hoc coupling to some
moduli. We leave a detailed investigation of string creation to a
future work \cite{morework}.

In this paper we have restricted attention to ``small" string
states, namely strings with oscillating modes only. It can be
expected that the phenomenon of string mode creation and its induced
effects on the equation of motion prevails for the long strings
which stretch with the expansion of the universe. Again, detailed
result will be presented in \cite{morework}.

We also expect the new term we discovered will have some effects for
cosmic strings at later times.

%%%%%%%%%%%%%%%%%%
\acknowledgments

This work was supported by grants from CNSF. M.L. and Y.S. would
like to thank KIAS, CQUEST as well as APCPT for their hospitality,
where some of this work was done.
%%%%%%%%%%%%%%%%%%

\section{Appendix}

To investigate the question of mode creation, we need to consider
the expectation value of the occupation number operator on a state
that does not change with time, that is, the eigenstates of the time
independent number operator. Because $\dot{A}^i_{\pm m}$ and
$\dot{\tilde{A}}^i_{\pm m}$ do not vanish, the corresponding number
operators $\hat{N}$ and $\hat{\tilde{N}}$ depend on time. Thus we
need to find a set of creation and annihilation operators $b^i_m$
and $\tilde{b}^i_m$, s.t. $\dot{b}^i_{\pm m}=\dot{\tilde{b}}^i_{\pm
m}=0$. $b^i_{\pm m}$ and $\tilde{b}^i_{\pm m}$ are linear
combinations of $A^i_{\pm m}$ and $\tilde{A}^i_{\pm m}$. As we have
mentioned under (\ref{chbassis}), the most general form of linear
transformation preserving the commutation relation is to set \ba&&
b^i_m=\cosh(\gamma'_m)e^{i\phi'_m+i\delta'_m}A^i_m+\sinh(\gamma'_m)
e^{i\phi'_m}\tilde{A}^i_{-m}\
,\\&&\tilde{b}^i_m=\cosh(\gamma'_m)e^{i\psi'_m+i\delta'_m}\tilde{A}^i_m+\sinh(\gamma'_m)
e^{i\psi'_m}A^i_{-m}\ .\ea Demanding that
\ba\dot{b}^i_m&&=[i(\dot{\phi}'_m+\dot{\delta}'_m)\cosh(\gamma'_m)e^{i\phi'_m+i\delta'_m}
+\dot{\gamma}'_m\sinh(\gamma'_m)e^{i\phi'_m+i\delta'_m}+c_m\sinh(\gamma'_m)e^{i\phi'_m}]A^i_m\\
\n&&+[i\dot{\phi}'_m\sinh(\gamma'_m)e^{i\phi'_m}
+\dot{\gamma}'_m\cosh(\gamma'_m)e^{i\phi'_m}+c^*_m\cosh(\gamma'_m)
e^{i\phi'_m+i\delta'_m}]\tilde{A}^i_{-m}=0\
,\\\dot{\tilde{b}}^i_m&&=[i(\dot{\psi}'_m+\dot{\delta}'_m)\cosh(\gamma'_m)e^{i\psi'_m+i\delta'_m}
+\dot{\gamma}'_m\sinh(\gamma'_m)e^{i\psi'_m+i\delta'_m}+c_m\sinh(\gamma'_m)e^{i\psi'_m}]\tilde{A}^i_m\\
\n&&+[i\dot{\psi}'_m\sinh(\gamma'_m)e^{i\psi'_m}
+\dot{\gamma}'_m\cosh(\gamma'_m)e^{i\psi'_m}+c^*_m\cosh(\gamma'_m)
e^{i\psi'_m+i\delta'_m}]A^i_{-m}=0\ .\ea Then we have
\ba&&\dot{\gamma}'_m+\Re(c_me^{-i\delta'_m})=0\
,\\&&\dot{\delta}'_m+[\tanh(\gamma'_m)+\coth(\gamma'_m)]\Im(c_me^{-i\delta'_m})=0\
.\ea Suppose that at time $t=t_0$, the two sets of operators are
identical, e.g. $b^i_m=A^i_m(t_0),\,\tilde{b}^i_m=\tilde{A}^i_m(t_0)
$. Then the Hilbert space spanned by eigenstates of $N'\equiv
b^i_{-m}b^i_m$ and $\tilde{N}'\equiv \tilde{b}^i_{-m}\tilde{b}^i_m$
is identical with that of $\hat{N}(t_0)$ and $\hat{N}(t_0)$. Thus we
have the initial condition $\gamma'_m=\delta'_m=0$. Denote $|t_o>$ a
state that does not change with time, then $<t_0|\hat{N}(t)|t_0>$
represents the change of the total mode number.
\be<t_0|\dot{N}+\dot{\tilde{N}}|t_0>=-\sum_{m,i}2mH\sinh(2\gamma'_m)\cos(\delta'_m)(N^i_m
+\tilde{N}^i_m+1)\ .\ee When $\vec{p}=0$, both the real part and the
imaginary part of $c_m$ is pure oscillating, so the average of
$\dot{\gamma}'_m$ and $\dot{\delta}'_m$ vanish, with the initial
condition, we will have $\gamma'_m\approx\delta'_m\approx0$. Then
$<t_0|\dot{N}|t_0>\approx0$.

\vfill
\eject
%%%%%%%%%%%%%%%%%%%%%%%%%%%%%%

%%%%%%%%%%%%%%%%%%%%%%%%%%%%%%
%%%%%%%%%%%%%%%%%%%%%%%%%%%%%%
\end{document}